\documentclass[aps,pra,twocolumn,showpacs,groupedaddress]{revtex4} 
\usepackage{graphicx}  
\usepackage{dcolumn}   
\usepackage{bm}       
\usepackage{amssymb}  
\usepackage{amsmath}
\usepackage{float}
\usepackage{enumerate}
\begin{document}
\title{Controlling of blow-up responses by a nonlinear $\cal{PT}$ symmetric coupling}
\author{S. Karthiga$^{1}$, V.K. Chandrasekar$^{2}$, M. Senthilvelan$^{1}$, M. Lakshmanan$^{1}$}
\address{$^1$Centre for Nonlinear Dynamics, School of Physics, Bharathidasan University, Tiruchirappalli - 620 024, Tamilnadu, India.\\
$^2$Centre for Nonlinear Science \& Engineering, School of Electrical \& Electronics Engineering, SASTRA University, Thanjavur -613 401, Tamilnadu, India.}
 \begin{abstract} 
\par  We investigate the dynamics of a coupled waveguide system with competing linear and nonlinear loss-gain profiles which can facilitate power saturation.  We show the usefulness of the model in achieving unidirectional beam propagation.  In this regard, the considered type of coupled waveguide system has two drawbacks, (i) difficulty in achieving perfect isolation of light in a waveguide and (ii) existence of blow-up type behavior for certain input power situations.  We here show a nonlinear $\cal{PT}$ symmetric coupling that helps to overcome these two drawbacks.  Such a nonlinear coupling has close connection with the phenomenon of stimulated Raman scattering.  In particular, we have elucidated the role of this nonlinear coupling using an integrable $\cal{PT}$ symmetric situation.  In particular, using the integrals of motion, we have reduced this coupled waveguide problem to the problem of dynamics of a particle in a potential.  With the latter picture, we have clearly illustrated the role of the considered nonlinear coupling.  The above $\cal{PT}$ symmetric case corresponds to a limiting form of a general equation describing the phenomenon of stimulated Raman scattering. We also point out the ability to transport light unidirectionally even in this general case. 
\end{abstract}
\maketitle
\section{Introduction}
\par Recently, $\cal{PT}$ symmetric systems have attracted wide interest in various fields such as optics \cite{r3,r4,r41,r42,r43,r44,r45}, plasmonics \cite{r5,r55}, quantum optics \cite{r6,r7}, Bose-Einstein condensation \cite{r8, r9}, acoustics \cite{acous}, and electronics \cite{r11,r12}.  Among them, the field of optics has gained interesting applications from $\cal{PT}$ symmetry, to mention a few, power oscillations \cite{r42, r43, re1}, unidirectional invisibility \cite{r43,re2,re3} and mode management in laser structures \cite{re4,re5,re6}.  Importantly in the construction of unidirectional photonic devices, the $\cal{PT}$ symmetric systems are found to be of considerable relevance \cite{nature, prl, pra, nature2,pra_sk2}.  Such undirectional transport of light is achieved in these systems when the $\cal{PT}$ symmetry is spontaneously broken.  As these $\cal{PT}$ symmetric systems are constructed by coupling a system with gain to a system with loss, in the symmetry broken region, independent of input power configurations,  these systems show a predominant transport of light through the waveguide with gain.  Such an ability to guide light unidirectionally indicates their applications over the construction of novel classes of integrated photonic devices such as optical diodes and optical isolators.  
\par In this connection, a difficulty that one can identify while applying the usual $\cal{PT}$ symmetric systems is with the intensity of light that is localized in the gain waveguide.  The intensity of light localized in the gain site is often quite high and further it varies with respect to the intensity of input light.  These two factors may turn out to be disadvantageous in practical situations.  One of the ways to overcome these difficulties is to introduce saturating effects in power through competiting linear and nonlinear loss-gain \cite{pra1, pre1} terms.   We call such loss-gain profile that helps to saturate power as saturating loss-gain profile.   Due to the presence of such saturating gain and loss, the output power in the system is maintained finite and it does not change with repsect to the input power configurations.  This type of stable finite power light transport is more useful from an application point of view \cite{srep, aac} and one can also find an interesting recent effort taken by Kominis {\it et al} to an asymmetric active coupler with unbalanced  linear loss-gain \cite{srep}.  
\par Although the above type of systems with saturating balanced loss-gain give controlled output power which is unaltered with the change in the input power configurations, these systems also have demerits.  The first one is with the inability to localize fully the light in a waveguide of the coupled system and the other is the existence of blow-up responses for certain input powers.  The existence of blow-up solutions in $\cal{PT}$ symmetric systems has also been reported earlier in \cite{stoke2} and it is obviously undesirable in practical situation.  To overcome the first problem, one may use self-trapping nonlinearity to localize light in one of the waveguides.  But considering the second demerit, we find that even self-trapping nonlinearity does not play useful role.   In the considered type of models, we find an interesting nonlinear $\cal{PT}$ symmetric coupling that is related to the phenomenon of stimulated Raman scattering, which has been found to be useful in overcoming these blow-up responses and also to help in the localization of light.  With an integrable $\cal{PT}$ symmetric situation, we have clearly demonstrated the usefulness of this nonlinear coupling in achieving a better unidirectional optical element.  However the $\cal{PT}$ symmetric situation is found to be a limiting case of a practically relevant and more general asymmetric coupled waveguide system including the effect of stimulated Raman scattering.   Even in the latter case, we show that we can achieve unidirectional transport of light and the nonlinear coupling is still useful in controlling blow-up and in localizing light. 
\par From a different perspective, the considered type of systems with saturating loss and gain can also be regarded as systems where a limit cycle oscillator is coupled to a system with inverted loss-gain profile (that is if the limit cycle oscillator has linear gain and nonlinear loss, the other system will have linear loss and nonlinear gain).  We have also studied the system from the dynamical point of view. 
\par To illustrate the above, we have prepared this article in the following manner.  In Sec. II, we have clearly presented the wave propagation property in a coupled waveguide system that is connected to the phenomenon of stimulated Raman scattering.  In Sec. III, we have studied the stability of symmetric and asymmetric modes of the $\cal{PT}$ symmetric case of the system.  In Sec. IV, we have presented the need of nonlinear $\cal{PT}$ symmetric coupling in this coupled waveguide system.  To more clearly demonstrate the role of nonlinear  coupling in controlling, we have used Sec. V and VI.  In Sec. V, we have shown the integrable nature of the system and in Sec. VI, we have used the integrable nature to reduce the coupled waveguide problem to a problem of particle in a potential well. With this particle in a potential well picture, we have clearly elucidated the dynamics of the coupled waveguide system in different parametric regions and demonstrated the role of considered nonlinear $\cal{PT}$ symmetric coupling in controlling the blow-up solutions.  In the earlier sections, we have illustrated the role of nonlinear coupling in a $\cal{PT}$ symmetric case while this may be treated as a limiting case of an experimentally realizable more general asymmetric coupled waveguide system.  As in the $\cal{PT}$ symmetric case, the ability to achieve unidirectional light transport and the role of nonlinear coupling related to the stimulated Raman scattering in controlling blow-up responses are illustrated for this asymmetric case in Sec. VII.  Finally, we have summarized our results in Sec. VIII. 
\section{\label{modeel}Model}
\par We here consider a coupled waveguide system in which the propagation of light with respect to the propagation distance $z$ is described by
\begin{eqnarray}
i \frac{d\phi_1}{dz}&=&\omega_1 \phi_1- i \gamma \phi_1+i \alpha |\phi_1|^2 \phi_1 -k \phi_2 -i \chi_1 |\phi_2|^2 \phi_1, \nonumber \\
i \frac{d\phi_2}{dz}&=&\omega_2 \phi_2+ i \gamma \phi_2- i \alpha |\phi_2|^2\phi_2-k \phi_1+i \chi_2 |\phi_1|^2 \phi_2,
\nonumber \\ 
\label{syst_0}
\end{eqnarray}
where, $\phi_1$ and $\phi_2$ are the complex amplitudes of electromagnetic field in the waveguides.  Throughout the manuscript, we have considered all the parameters given in (\ref{syst_0}) to be positive.   $\omega_1$ and $\omega_2$ correspond to the propagation constants and $k$ corresponds to the coupling strength which arises due to the interaction of evanascent fields. Further in (\ref{syst_0}), $\gamma$ is a linear loss-gain strength parameter and $\alpha$ is the nonlinear loss-gain strength which enables saturation of power in the coupled waveguide system. This nonlinear loss-gain can arise in a system when the imaginary part of their refractive index profile becomes nonlinear.  In (\ref{syst_0}), we have also introduced another interesting nonlinear coupling with strengths $\chi_1$ and $\chi_2$, where $\chi_1=\frac{\omega_1}{\omega_2}\chi$ and $\chi_2=\chi$. { The above type of nonlinear coupling arises due to the presence of complex components in the third order nonlinear susceptibility tensor.  The form of nonlinear coupling given in (\ref{syst_0}) is similar to the form of coupling observed between the complex amplitudes of pump and the Stokes modes in the stimulated Raman scattering process \cite{agar,stim}.  Thus the above mentioned nonlinear coupling can be achieved by sending a higher frequency pump signal in one waveguide and lower frequency Stokes signal in other waveguide so that we consider $\omega_1 > \omega_2$.  As it is known earlier, in order to achieve stimulated Raman scattering not only the light with pump frequency should be given as input but also the mode with Stokes frequency should be given as input.  To show the interesting characteristics of this nonlinear coupling in a simple way, we wish to consider an interesting $\cal{PT}$-symmetric case $\omega_1 =\omega_2$ (as the difference between the frequencies of pump and Stokes modes is small) in-detail,
\begin{eqnarray}
i \frac{d\phi_1}{dz}&=&\omega \phi_1- i \gamma \phi_1+i \alpha |\phi_1|^2 \phi_1 -k \phi_2 -i \chi |\phi_2|^2 \phi_1, \nonumber \\
i \frac{d\phi_2}{dz}&=&\omega \phi_2+ i \gamma \phi_2- i \alpha |\phi_2|^2\phi_2-k \phi_1+i \chi |\phi_1|^2 \phi_2.
\nonumber \\ 
\label{syst}
\end{eqnarray}
In particular, we show that the above limiting case is integrable.  This nature facilitates us to illustrate that the nonlinear coupling is useful in arresting the blow-up responses and in enhancing unidirectional transport of light as demonstrated through analytical and numerical means.  We further demonstrate that the above characteristics extend even to the case $\omega_1 \neq \omega_2$ (in Sec. VII).

\par Considering Eq. (\ref{syst}), we can check that this coupled waveguide system is invariant under the combined operation of parity and time reversal symmetries defined by $\phi_2 \rightarrow -\phi_1$, $\phi_1 \rightarrow -\phi_2$, $i \rightarrow -i$ and $z \rightarrow -z$.  But in the absence of coupling, the system is not $\cal{PT}$ symmetric and the individual systems are found to have unbalanced loss-gain profile.  While isolated ($k=0$, $\chi=0$), one can find that the first waveguide has linear loss ($\gamma>0$) and nonlinear gain ($\alpha>0$).  So the amplification in the power introduced by nonlinear gain ($\alpha>0$) is opposed by the linear loss ($\gamma>0$) due to which the input power dampens to zero as $z \rightarrow \infty$.   Note that in the uncoupled situation ($k=\chi=0$), the equation describing the propagation of light in the second waveguide is of  similar form as that of the well known Stuart-Landau oscillator.  We observe the existence of linear gain ($\gamma>0$) and nonlinear loss ($\alpha>0$) in the second waveguide.  Here the linear gain makes the power to grow and as soon as the power builds up the intensity dependent nonlinear loss ($\alpha>0$) comes into action and power will be saturated to a constant value.  Due to the above fact, in this case, one observes limit cycle oscillations in $\phi_1$ as $z \rightarrow \infty$ with constant amplitude/power $|\phi_1|^2=\frac{\gamma}{\alpha}$.  
\par Even under the coupled situation, the competing linear and nonlinear loss-gain will not allow the power to grow up for very high values (excluding blow-up regions) which we will show in the following section.  Due to this advantage, we expect that this type of systems may be more useful in the application point of view.   We discuss how far this system can be used for the construction of unidirectional devices and the usefulness of nonlinear $\cal{PT}$ symmetric coupling in achieving unidirectional light propagtion in coupled waveguides.   In the absence of this nonlinear coupling, the dynamics of the system has been studied in \cite{pra1, pre1}.  However achieving unidirectional transport of light and controlling blow-up responses in these systems have not been discussed in these works.  Also the interesting role of nonlinear coupling in this type of systems has not yet been studied in the literature and here we show its role using an interesting particle in a potential picture.
\section{Different modes and their stability}
\par First considering the linear form of (\ref{syst}), the linear symmetric mode is found to be stable for $k\geq \gamma$ and it becomes unstable with the increase of $\gamma$ \cite{pra}.  Let us now seek the nonlinear stationary modes of the system (\ref{syst}) by considering 
\begin{eqnarray}
\phi_1= R_1 e^{-i \tilde{\omega} z+i \theta_1} \;\; \mathrm{and}\;\; \phi_2=R_2 e^{-i \tilde{\omega} z+i \theta_2}.
\label{ph1ph2}
\end{eqnarray}
 Subtituting the above in (\ref{syst}), we obtain
\begin{eqnarray}
\dot{R_1}&=&-\gamma R_1+\alpha R_1^3+k R_2 \sin \delta-\chi R_2^2 R_1, \nonumber \\
\dot{R_2}&=&\gamma R_2-\alpha R_2^3-k R_1 \sin \delta+\chi R_1^2 R_2,  \nonumber \\
\dot{\delta}&=&-\frac{k(R_1^2-R_2^2)}{R_1 R_2} \cos \delta,
\label{rtheta}
\end{eqnarray}
where the overdot represents differentiation with respect to $z$, and $\delta=\theta_1-\theta_2$. 
From the above, we find that the system admits symmetric mode with, 
\begin{eqnarray}
R_1^*=R_2^*= R^*, \quad \sin \delta^* = \frac{\gamma-(\alpha-\chi) {R^*}^2}{k}
\label{s_mode}
\end{eqnarray}
Note that in the above expression $R^*$ can take any positive value and it finds a restriction putforth by the expression of $\sin \delta$.  As the values of $|\sin \delta| \leq 1$, $R^*$ can lie only in the range $0\leq \frac{k-\gamma}{(\chi-\alpha)} \leq {R^*}^2\leq \frac{k+\gamma}{(\chi-\alpha)}$ (Note that $R^*$ is positive valued).  Depending on the input power configurations, the system will take one of the values of $R^*$ in this range.  By perturbing the above symmetric modes, we have obtained the eigenvalues characterizing the stability of the mode.  
\begin{figure*}
\includegraphics[width=0.8\linewidth]{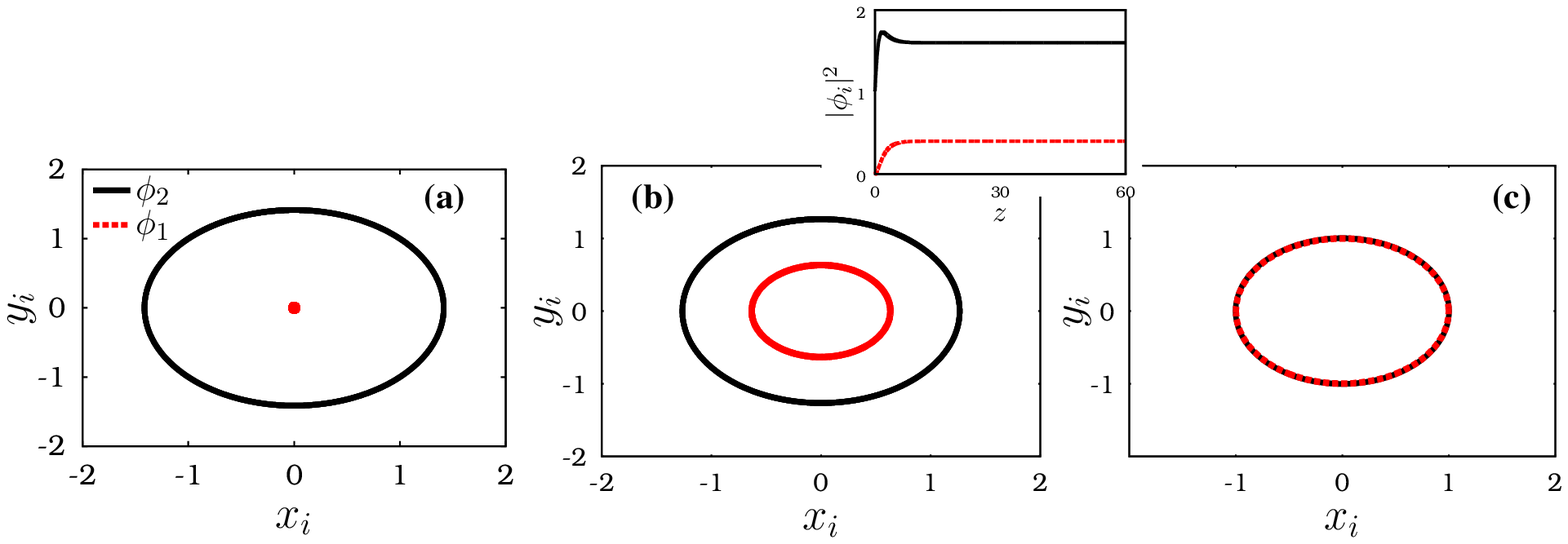}
\caption{(color online) Limit cycle behaviours of $\phi_1$ and $\phi_2$: Figures are plotted for $\gamma=1.0$, $\alpha=0.5$, and $\chi=0.0$.  Fig. (a), (b) and (c) are plotted for $k=0.0$, $0.4$ and $0.5$ respectively.  Here the dynamics of the system is represented in terms of the variables $x_j$ and $y_j$, where $\phi_j=x_j +i y_j$, and $\phi_1$ is represented by dashed red line and $\phi_2$ is represented by continuous black line. The inset in the figure (b) shows the variation in $|\phi_1|^2$ and $|\phi_2^2|$ with respect to $z$.  }
\label{iso}
\end{figure*}
The eigenvalues corresponding to the symmetric mode (\ref{s_mode}) are
\begin{small}
\begin{eqnarray}
\lambda=0, \pm 2\sqrt{\gamma^2-k^2-3 R_2^2 \gamma(\alpha-\chi)+R_2^4 (3 \alpha^2-4 \alpha \chi+\chi^2)} 
\label{syn_eig}
\end{eqnarray}
\end{small}
Whenever the term inside the square root in (\ref{syn_eig}) is negative, the symmetric mode becomes neutrally stable.  Such a neutrally stable nature indicates that the periodic solution assumed in (\ref{ph1ph2}) is modulated by a slowly varying amplitude so that we observe quasi-periodic oscillations in $\phi_i$.  While analyzing the stability of the symmetric mode, for $\alpha\geq \chi$ we find that the symmetric mode in (\ref{s_mode}) is found to be unstable for all possible values of $R^*$ in the parametric range $0<k<\frac{\gamma(\alpha+\chi)}{2 \alpha}$ (as $k \in {\bf R^+}$) and in the other paramteric ranges the symmetric mode is found to be stable for certain range of values of $R^*$.   Similarly for $\alpha<\chi$, we find that the symmetric mode is found to be unstable in the range $0<k< \gamma$ (as $k \in {\bf R^+}$) for all possible values of $R^*$.
\par   The set of amplitude and phase equations in (\ref{rtheta}) also suggest the existence of two asymmetric modes with $R_1^2+R_2^2=\frac{\gamma}{\alpha}$ and they are of the form
\begin{eqnarray}
\mathrm{(i)} \;\; R_1^*=\frac{1}{\sqrt{2}}\sqrt{\frac{\gamma}{\alpha}-\Delta}, \quad R_2^*=\frac{1}{\sqrt{2}}\sqrt{\frac{\gamma}{\alpha}+\Delta},
\label{as_mod1}
\end{eqnarray} 
and 
\begin{eqnarray}
\mathrm{(ii)} \;\; R_1^*=\frac{1}{\sqrt{2}}\sqrt{\frac{\gamma}{\alpha}+\Delta}, \quad R_2^*=\frac{1}{\sqrt{2}}\sqrt{\frac{\gamma}{\alpha}-\Delta},
\label{as_mod2}
\end{eqnarray} 
alongwith
\begin{eqnarray}
\delta=\frac{(2n+1) \pi}{2},\; n=0,1,2,..., 
\label{deld}
\end{eqnarray}
where
\begin{eqnarray}
\Delta=\sqrt{\frac{\gamma^2}{\alpha^2}-\frac{4k^2}{(\alpha+\chi)^2}}, \;\; 
\label{deldel}
 \end{eqnarray}
While studying the stability properties of the above modes, we observed that the mode (\ref{as_mod2}) is unstable for all values of the parameters and for all possible values of $\delta$ given in (\ref{deldel}) but the mode (\ref{as_mod1}) is unstable only when $\delta=\frac{(2n+1)\pi}{2}$, where $n=1,3,5,...$.  For $\delta=\frac{(2n+1)\pi}{2}$, where $n=0,2,4,...$, the eigenvalues corresponding to the mode (\ref{as_mod1}) are
\begin{small}
\begin{eqnarray}
\lambda=\frac{\left(-(3\alpha+\chi)\pm \sqrt{(3\alpha+\chi)^2-8 \alpha (\alpha+\chi) }\right) \Delta}{2},-(\alpha+\chi)\Delta. \quad
\end{eqnarray}
\end{small}
These eigenvalues correspond to the fact that the considered asymmetric mode is stable as long as it exists, that is
\begin{eqnarray}
\Delta>0 \;\; \mathrm{or} \;\; \frac{\gamma^2}{\alpha^2}-\frac{4k^2}{(\alpha+\chi)^2}>0 \; \; \Rightarrow \; \; k<\frac{\gamma(\alpha+\chi)}{2 \alpha}.
\label{fixcri}
\end{eqnarray}
This establishes that this asymmetric mode is stable for lower values of $k$.  Whenever the asymmetric modes are stable (that is in the region $k<\frac{\gamma(\alpha+\chi)}{2 \alpha}$), the $\cal{PT}$ symmetry nature of the system is spontaneously broken.  This is because of the reason that in the asymmetric mode propagation, the total power is not equally distributed among the first and second waveguides as evident from $R_1^* \neq R_2^*$ in the asymmetric modes.  The light is found to be more localized in the second waveguide than in the first waveguide and the intensity of light localized in the second waveguide or in the first waveguide is found to be finite.     Whenever the asymmetric mode is unstable or symmetric mode is stable, the symmetry is said to be preserved. 
\par Now looking at the form of the asymmetric mode given in (\ref{as_mod1}), we find that the amplitude of the mode depends only on the system parameters and coupling parameters, and it does not depend on the input power configurations.  Such a form of asymmetric mode indicates that there exists an isolated limit cycle type periodic attractor in $\phi_1$ and $\phi_2$  (see Eq.(\ref{ph1ph2})) with amplitudes $R_1^*$ and $R_2^*$ (as given in Eq. (\ref{as_mod1})).  Thus for all initial conditions near this attractor, we observe that the powers $|\phi_1|^2$ and $|\phi_2|^2$ tend to constant values ${R_1^*}^2$ and ${R_2^*}^2$ given in Eq. (\ref{as_mod1}). 

\begin{figure*}[htb!]
\includegraphics[width=0.8\linewidth]{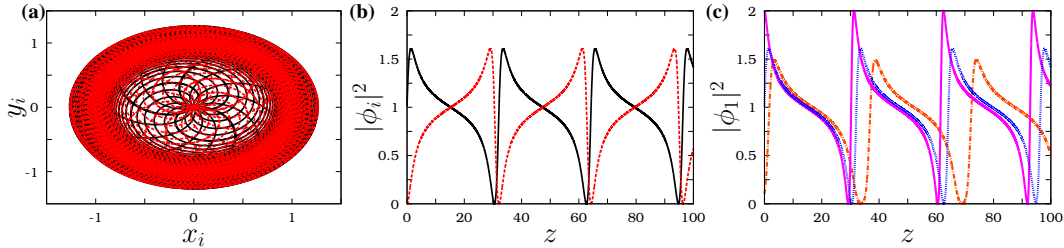}
\caption{(color online) Dynamics in the unbroken $\cal{PT}$ region: The figures are plotted for $\gamma=1.0$, $\alpha=0.5$, $\chi=0.0$ and $k=0.51$.  Fig. (a) is plotted for the initial condition $\phi_1=1.0$ and $\phi_2=0.0$.  From the figure, we find that the oscillations become quasi-periodic (where $\phi_j=x_j +i y_j$) and their corresponding power oscillations are given in Fig. (b). $\phi_1$ is represented by dashed red line and $\phi_2$ is represented by continuous black line in the Figures (a) and (b). Fig. (b) shows the equal distribution of power among the two waveguides so that the symmetry is now unbroken. By considering different initial conditions $\phi_1=\sqrt{0.1},1.0$ and $\sqrt{2.0}$ with $\phi_2=0.0$, we have plotted variation of $|\phi_1|^2$ in Fig. (c) and the existence of non-isolated power oscillations are shown in the figure. }
\label{quasi}
\end{figure*}
\par In the uncoupled case $k=\chi=0$, we find $R_1^*=0$ and $R_2^*=\sqrt{\frac{\gamma}{\alpha}}$ and so one observes that $\phi_1 \rightarrow 0$ and $\phi_2$ executes oscillation with amplitude $R_2^*$ as shown in Fig. \ref{iso}.  Thus there exists constant power propagation in the second waveguide (where $|\phi_1|^2=\frac{\gamma}{\alpha}$ as $z \rightarrow \infty$) under the uncoupled situation whereas the input power will be completely dissipated in the first waveguide.  Now introducing the linear coupling $k$ as $k=0.4$ (we here first study the role of linear coupling alone where $\chi=0.0$ and the role of nonlinear coupling is demonstrated in the next section), from (\ref{as_mod1}) we observe both $R_1^*, R_2^* \neq 0$ so that limit cycle type oscillations are observed in both $\phi_1$ and $\phi_2$ as shown in Fig. \ref{iso}(b).  Now the light propagates in both the waveguides where the intensities in the two waveguides approach constant values as $z \rightarrow \infty$  as shown in the inset of Fig. \ref{iso}(b).  Eq. (\ref{as_mod1}) also implies that the value of the total power $p=|\phi_1|^2+|\phi_2|^2$ remains invariant with respect to the values of $k$ (and also $\chi$).  The increase in the coupling $k$ gives rise to the growth of limit cycle oscillations in $\phi_1$ and suppression in the amplitude of the limit cycle oscillations in $\phi_2$ or increase in the value of $|\phi_1|^2$ and decrease in $|\phi_2|^2$ as represented by (\ref{as_mod1}).  At $k=\frac{\gamma(\alpha+\chi)}{2 \alpha}$, the amplitudes of both $\phi_1$ and $\phi_2$ become equal as shown in Fig. \ref{iso}(c).  The intensities of light in both the waveguides are equal now.  Such equalization in powers, $|\phi_1|^2$ and $|\phi_2|^2$, denotes the unbroken nature of the $\cal{PT}$ symmetry. Beyond this, an increase in $k$ will not lead to an increase in the amplitude of $\phi_1$ with corresponding decrease in the amplitude of $\phi_2$.  
\par On increasing the value of $k$ beyond $\frac{\gamma(\alpha+\chi)}{2 \alpha}$, we find that the amplitudes $\phi_1$ and $\phi_2$ remain equal and we observe quasi-periodic like oscillations in $\phi_1$ and $\phi_2$.  Fig. \ref{quasi}(a) shows the existence of quasi-periodic oscillations in the system for $k=0.51>\frac{\gamma(\alpha+\chi)}{2 \alpha}$ (where $\gamma=1.0$, $\alpha=0.5$ and $\chi=0.0$). This shows the correctness of the results obtained from stability of symmetric modes where we have seen that for $\alpha>\chi$, the symmetric mode starts to become stable for $k>\frac{\gamma(\alpha+\chi)}{2 \alpha}$.  Fig. \ref{quasi}(b) shows the power in the two waveguides for the same parametric values and we can find that the waveguides carry equal power or we can find the propagation of symmetric mode.    The symmetry is said to be unbroken now.  In this unbroken $\cal{PT}$ region, the oscillations are not isolated and are found to be varying with respect to the initial conditions.  To show the above, we have plotted Fig. \ref{quasi}(c) for different initial conditions and see that the power oscillations in the system varies with respect to the input power configurations.  The value of power is also restricted as mentioned earlier in (\ref{s_mode}) where the value of $R$ can take any values in the range ${R^*}^2\geq \frac{k-\gamma}{(\chi-\alpha)}$ and ${R^*}^2\leq \frac{k+\gamma}{(\chi-\alpha)}$.

{\par   Thus it is clear from the above discussion that the oscillations (observed in $\phi_i$) corresponding to asymmetric modes are isolated whereas in the symmetric modes, there exists non-isolated set of oscillations. These isolated limit cycles are found to be stable for lower coupling strength $k$ and the non-isolated symmetric orbits are found to be stabilized by the increase of $k$.   The reason is that in the uncoupled case, the individual systems have unbalanced loss-gain profiles, so that one of the systems has linear loss and nonlinear gain, the other system has linear gain and nonlinear loss.  Due to the unbalanced nature of the loss-gain profile, we observe that there are only isolated closed paths in $\phi_i$.  But in the presence of balanced loss-gain, the orbits need not be isolated \cite{pra_sk1}.  Thus with the introduction of coupling, the linear gain and nonlinear loss given in $\phi_1$ try to balance with the linear loss and nonlinear gain present in $\phi_2$.  Thus an increase in $k$ stabilizes many closed paths so that we observe non-isolated orbits in the unbroken $\cal{PT}$ region.}  
\par  So far in the above discussion, we have studied the role of coupling $k$ in the system (\ref{syst}).  Then, the interesting roles of nonlinear $\cal{PT}$ symmetric coupling and its application are demonstrated in the next section by exploiting the integrable nature of the considered system.
\section{Role of nonlinear $\cal{PT}$ symmetric coupling} 
\par{  Before studying the role of nonlinear $\cal{PT}$ symmetric coupling in the system, we here first review whether the observed dynamics is found to be useful from an application point of view.  Considering a $\cal{PT}$ symmetric system, the literature reveals that the dynamics in the broken $\cal{PT}$ phase is more interesting than the dynamics in the unbroken $\cal{PT}$ phase \cite{nature, prl, pra}.  The reason is that in the broken $\cal{PT}$ phase, the evolution of light can be unidirectional and further the light is found to be localized in the gain site independent of the input power configuration.  The problem with such $\cal{PT}$ symmetric systems (without saturating nonlinear loss-gain) is that the intensity of light localized in the gain site can be quite high. Such a high intense light output may not be favorable in practical situations.  
\par Considering our case, the inset in Fig. \ref{iso}(b) shows that unlike the usual $\cal{PT}$ symmetric systems without saturating gain and loss, here the intensity of light is not much higher in the broken region.  Importantly, in the asymptotic limit the intensities of light in the two waveguides do not vary with respect to input power configurations. These two qualities hinders our aim to achieve unidirectional light transport in this type of $\cal{PT}$ symmetric systems.
\begin{figure}[htb!]
\includegraphics[width=0.6\linewidth]{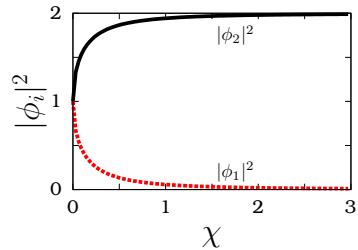}
\caption{(color online) Figure shows the values of power in the first and second waveguides for different values of $\chi$ and it is plotted for $k=0.5$, $\gamma=1.0$ and $\alpha=0.5$.  From the figure, we can observe the isolation of light in the second waveguide with increase of $\chi$. }
\label{chi}
\end{figure}   
\par Additionally, one can find from Fig. \ref{iso}(b) (and also from Eq. (\ref{as_mod1})) that although the intensity of light propagating in the second waveguide is higher than the intensity of light propagating in the first waveguide, light is not completely localized in one of the waveguides. So in order to the make use of these systems for unidirectional transport devices, we are in need to localize the light in one of the waveguides only.  To tackle the above problem, the nonlinear $\cal{PT}$ symmetric coupling considered in (\ref{syst}) is more useful.  In Fig. \ref{chi}, we have chosen $k=0.5$, $\gamma=1.0$ and $\alpha=0.5$ (where the parametric values correspond to the broken $\cal{PT}$ region) and plotted the power in the first and second waveguides for different values of $\chi$.  The figure shows that an increase in the value of $\chi$ makes the total power ($p=\frac{\gamma}{\alpha}=2$) to be localized in the second waveguide and the power in the first waveguide tends to zero. If one introduces nonlinearity such as self-trapping/self-focussing type, one can achieve complete localization of light even for very low values of nonlinear $\cal{PT}$ symmetric coupling strength.  This type of complete localization of light in the second waveguide allows the possibility to use this system in practical situations. It is also important to note that the input power independent beam dynamics in these system indicates that we can get the same unidirectional light evolution even for low input powers.  This provides advantage in low power applications.

\par But considering higher input powers, the studies on their dynamics show that these systems show blow-up response (we also wish to note that even the other $\cal{PT}$ symmetric systems without saturating loss-gain show blow-up solutions for high input powers).  If one controls such blow-up responses in the system, then the system can be used for a very wide range of input powers.  For this purpose, we have tested whether self-trapping nonlinearity can control such blow-up solutions.   Although the self-trapping nonlinearity helps in localizing the light in one of the waveguides, it does not play any role in controlling blow-up responses.   But the considered nonlinear coupling plays a very interesting role in controlling such blow-up responses as well.  To illustrate the above, in the following sections, we exploit the integrable nature of the system (\ref{syst}) in terms of the so called Stokes variables \cite{aac,stoke2} and transform the considered coupled waveguides problem to the problem of particle in a potential.  With the latter picture, we illustrate the existence of blow-up solutions in the absence of $\chi$ for certain input powers and the role of $\chi$ in controlling such blow-up regions.

\section{Integrability of the system - Stokes variable dynamics}
\par The dynamics of the system (\ref{syst}) can be studied in terms of certain real variables, namely the Stokes variables \cite{pra,pre1}.  These Stokes variables can be defined as
\begin{eqnarray}
p&=& |\phi_1|^2+|\phi_2|^2, \quad
s_0= |\phi_1|^2-|\phi_2|^2, \nonumber \\
s_1&=& \phi_1 \phi_2^*+\phi_1^* \phi_2, \quad  s_2= i(\phi_1 \phi_2^*-\phi_1^* \phi_2),
\label{stok}
\end{eqnarray}
where $p$ denotes the total power in the system and $s_0$ denotes the difference in the intensities of first and second waveguides. One can also find that there exists a relation among these four Stokes variables as
\begin{eqnarray}
p^2=s_0^2+s_1^2+s_2^2.
\label{prel}
\end{eqnarray}
The coupled waveguide equation given in (\ref{syst}) in terms of these Stokes variables can be written as
\begin{eqnarray}
&&\dot{p}=-2(\gamma-\alpha p) s_0, \label{pd}  \\
&&\dot{s_0}=-2(\gamma-\alpha p) p-\left[(\alpha+\chi) s_1\right] s_1 \nonumber \\
&& \qquad \qquad \qquad -\left[(\alpha+\chi) s_2+2 k\right] s_2, \label{s0d} \\
&& \dot{s_1}=\left[(\alpha+\chi)s_1\right] s_0, \label{s1d} \\
&& \dot{s_2}=\left[(\alpha+\chi) s_2+2 k\right] s_0. \label{s2d}
\end{eqnarray}
From Eqs. (\ref{pd}) and (\ref{s1d}), we find that 
\begin{eqnarray}
 \frac{\dot{s_1}}{(\alpha+\chi)s_1}=\frac{-\dot{p}}{2(\gamma-\alpha p)},
\end{eqnarray}
From the above, one can directly find that  
\begin{eqnarray}
s_1=I_1 (\gamma-\alpha p)^{\frac{(\alpha+\chi)}{2 \alpha}},
\label{exp_s11}
\end{eqnarray}
where $I_1$ is an integral of motion.  It can be fixed from the initial conditons as
\begin{eqnarray}
I_1=\frac{ s_1(0)}{(\gamma-\alpha p(0))^{\frac{\alpha+\chi}{2 \alpha}}},
\label{exp_i11}
\end{eqnarray}
where $s_1(0)=|\phi_1|^2(0)-|\phi_2|^2(0)$ and $p(0)=|\phi_1|^2(0)+|\phi_2|^2(0)$.
The expressions (\ref{exp_s11}) and (\ref{exp_i11}) are valid only if $p(0)< \frac{\gamma}{\alpha}$.  For $p(0)>\frac{\gamma}{\alpha}$, $I_1$ obtained from (\ref{exp_i11}) may not be real.  Thus for $p(0)> \frac{\gamma}{\alpha}$, we have
\begin{eqnarray}
s_1=I_1 (\alpha p-\gamma)^{\frac{(\alpha+\chi)}{2 \alpha}},
\label{s11p}
\end{eqnarray}
where now
\begin{eqnarray}
I_1=\frac{ s_1(0)}{(\alpha p(0)-\gamma)^{\frac{\alpha+\chi}{2 \alpha}}}.
\end{eqnarray}
In a similar way, from Eqs. (\ref{pd}) and (\ref{s2d}), one can find that for $p(0)<\frac{\gamma}{\alpha}$,
\begin{eqnarray}
s_2=\frac{I_2 (\gamma-\alpha p)^{\frac{(\alpha+\chi)}{2 \alpha}}-2k }{(\alpha+\chi)},
\label{s21p}
\end{eqnarray}
where
\begin{eqnarray}
I_2=\frac{(\alpha+\chi) s_2(0) +2 k}{(\gamma-\alpha p(0))^{\frac{(\alpha+\chi)}{2 \alpha}}},
\label{exp_i2}
\end{eqnarray}
and for $p(0)>\frac{\gamma}{\alpha}$,
\begin{eqnarray}
s_2=\frac{I_2 (\alpha p-\gamma)^{\frac{(\alpha+\chi)}{2 \alpha}}-2k }{(\alpha+\chi)},
\label{s22p}
\end{eqnarray}
where now
\begin{eqnarray}
I_2=\frac{(\alpha+\chi) s_2(0) +2 k}{(\alpha p(0)-\gamma)^{\frac{(\alpha+\chi)}{2 \alpha}}}.
\label{exp_i2}
\end{eqnarray}
The existence of two integrals of motion ($I_1$ and $I_2$) associated with the system (\ref{pd})-(\ref{s2d}) of four coupled nonlinear ordinary differential equations of autonomous type  ensures that the system is integrable.  With these integrals $I_1$ and $I_2$, we can reduce the set of equations given in Eqs. (\ref{pd}) - (\ref{s2d}) to a first order differential equation as
\begin{eqnarray}
\dot{p}^2= 4(\gamma-\alpha p)^2 ({p^2-s_1^2(p)-s_2^2(p)})
\label{first}
\end{eqnarray}
where we have to substitute (\ref{pd}) in (\ref{prel}), and $s_1$ and $s_2$ take the forms given in (\ref{exp_s11}), (\ref{s11p}), (\ref{s21p}) and (\ref{s22p}).  The above equation (\ref{first}) can be solved in principle by quadratures,  though explicit solution can be given only for special choices of $s_1(p)$ and $s_2(p)$ (or parameters).   Thus we proceed to analyze the following in yet another way to understand the dynamics from this reduced equations, the details of which are given in the next section. 
\begin{figure}
\includegraphics[width=1.1\linewidth]{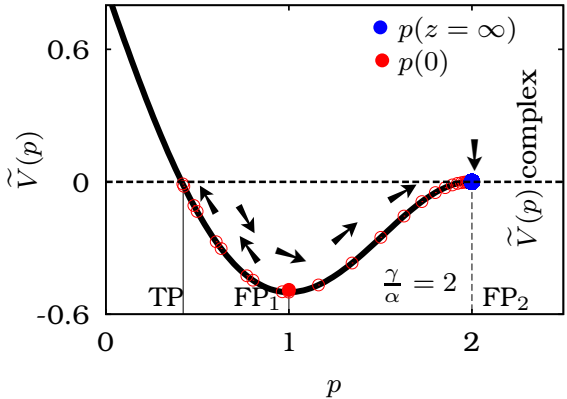}
\caption{(color online) Dynamics in the broken $\cal{PT}$ symmetric region for $\chi=0.0$: Figure shows the potential curve of $\widetilde{V}(p)$ for $\gamma=1.0$, $\alpha=0.5$, $k=0.4$, and $\chi=0.0$.  It is obtained for the initial condition $\phi_1=1.0+0.0 i$ and $\phi_2=0.0$. The empty circles along the $\widetilde{V}(p)$ curve represent the dynamics of $p$ with respect to $z$.  }
\label{fig1}
\end{figure}
\section{Understanding from particle dynamics in a potential}
\par Note that Eq. (\ref{first}) can be written in an interesting form as
\begin{eqnarray}
\frac{\dot{p}^2}{2}+V(p,E)=0,
\label{hameq}
\end{eqnarray}
where
\begin{small}
\begin{align}
V(p,E)=\widetilde{V}(p)=V(p)-E=2 (\gamma-\alpha p)^2\big[s_1^2(p)+s_2^2(p)-p^2\big]. \quad
\label{potent}
\end{align}
\end{small}
In the above, $s_1(p)$ and $s_2(p)$ are found to take appropriate forms, (depending on the value of $p(0)$) given in (\ref{exp_s11}), (\ref{s11p}), (\ref{s21p}) and (\ref{s22p}).
Equation (\ref{hameq}) looks similar in form as that of the expression of total energy of a particle in the potential $V(p)$ with total energy $E$ which can be related to the integrals $I_1$ and $I_2$.  Thus the considered problem now interestingly gets transformed to the problem of studying dynamics of a particle in the potential $V(p)$.  In the latter picture, the position of the particle is represented by $p$, while $\dot{p}$ represents the velocity of the particle in the potential and the independent variable $z$ is considered here as the time variable.   The initial condition of the problem is fixed from the initial conditions of the coupled waveguide problem.  For example, the initial position of the particle is $p(0)=|\phi_1(0)|^2+|\phi_2(0)|^2$ and $\dot{p}(0)=-(\gamma-\alpha p(0)) s_0(0)$ (following Eq. (\ref{pd})).
\par An understanding on the form of the potential, the maxima and minima of the potential and turning points of oscillatory solutions all of which can serve well for the understanding of the considered problem.  In a similar way, by exploiting the structure of the potential, we here illustrate the dynamics observed in the coupled waveguide system and the role of nonlinear $\cal{PT}$ symmetric coupling in controlling blow-up regions. 
\begin{figure*}
\includegraphics[width=1.05\linewidth]{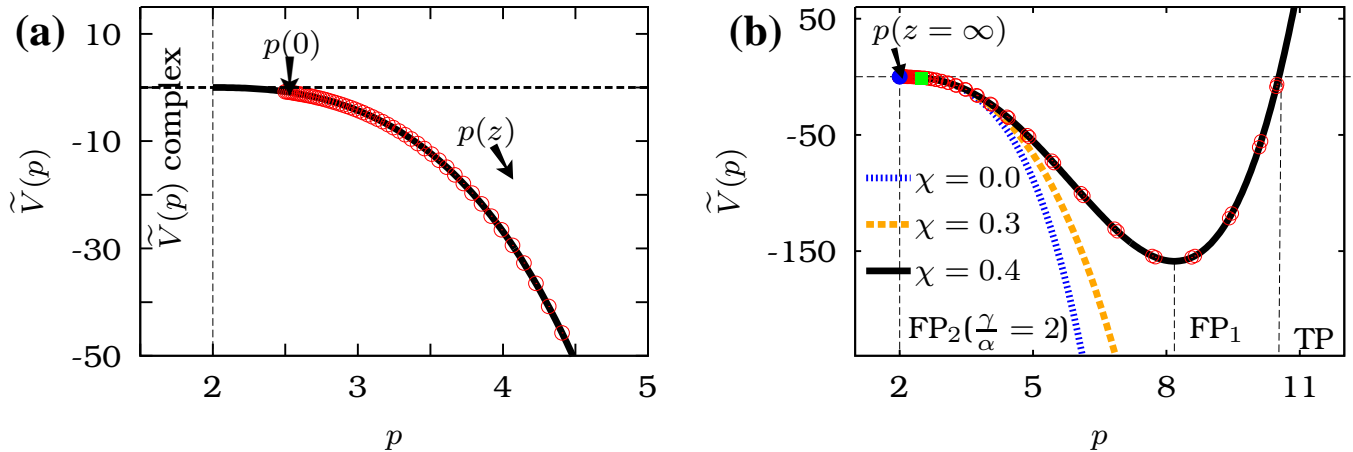}
\caption{(color online) Blow-up responses in the broken $\cal{PT}$ symmetric region: Fig. (a) is plotted for the same values of paramters as that of Fig. \ref{fig1} but for initial conditions $\phi_1=\sqrt{2.5}$ and $\phi_2=0.0$ and it also shows the existence of blow-up solution for certain input power configurations.  Fig. (b) shows the curves of $\widetilde{V}(p)$ for different values of $\chi$, namely, $0.0,0.3$ and $0.4$, for the same initial conditions and parametric values as considered in Fig. (a). }
\label{fig2}
\end{figure*} 
\par Considering the potential $\widetilde{V}(p)$, the steady states of the system can be obtained from $V'(p^*)=0$ as
\begin{eqnarray}
 p^*=\frac{\gamma}{\alpha},   
\label{fpp1}
\end{eqnarray}
and the ones satisfying
\begin{eqnarray}
-\alpha F(p^*)+  (\gamma-\alpha p^*) \frac{F'(p^*)}{2}=0, \quad
\label{com_fix}
\end{eqnarray} 
where $F(p^*)=s_1^2(p^*)+s_2^2(p^*)-{p^*}^2$.  To know whether these steady states correspond to maxima or minima of the potential, we have to check whether $V''(p^*)<0$ or $V''(p^*)>0$ at these fixed points. Considering the steady state $p^*=\gamma/\alpha$, 
\begin{eqnarray}
V''(p^*)=-4 \alpha^2 \Delta^2.
\label{stabi1}
\end{eqnarray}
Whenever $\Delta^2>0$, $V''(p^*)<0$, the state $p^*=\frac{\gamma}{\alpha}$ is a maximum, otherwise it is a minimum to the potential.  In other words, for lower values of $k$, $k<\frac{\gamma(\alpha+\chi)}{2 \alpha}$ the steady state is found to be maximum and for values of $k>\frac{\gamma(\alpha+\chi)}{2 \alpha}$ it is found to be a minimum.  The other useful quantity are the turning points which are obtained from $\widetilde{V}(p)=0$.  In our case, they are found to be
\begin{eqnarray}
p_{t}=\frac{\gamma}{\alpha},
\label{tp1}
\end{eqnarray} 
and  the ones satisfying the condition
\begin{eqnarray}
s_1^2(p_{t})+s_2^2(p_{t})-p_{t}^2=0.
\label{tpp}
\end{eqnarray}
Note that the turning point given in (\ref{tp1}) is not only a turning point but also a fixed point (see (\ref{fpp1})). 
\subsection{Dynamics in the broken $\cal{PT}$ region}
\subsubsection{Case: $\chi=0$}
\par Now with this new picture, let us analyze the dynamics of the coupled waveguide system (\ref{syst}).   For the purpose, we first consider a lower value of $k$, $k=0.4$, and study the dynamics of the system for lower input powers.  Next, for the same value of $k$, we show the existence of blow-up solutions for higher input powers and then also consider the role of $\chi$ in controlling these blow-up solutions.  First considering the initial condition $\phi_1(0)=1.0+0.0 i$ and $\phi_2(0)=0.0$, we have plotted the potential (\ref{potent}) for $\gamma=1.0$, $\alpha=0.5$ and $\chi=0.0$ in Fig. \ref{fig1}.  The dynamics of $p$ in the coupled waveguide problem is shown by empty circles along the potential in Fig. \ref{fig1}.  Here, we try to understand the dynamics as if a particle moves along the potential.  We try to find the fixed points of the system.  It is obvious that one of the steady states of the system is $p^*=\frac{\gamma}{\alpha}$ and it is denoted by FP$_2$ in Fig. \ref{fig1}. The nature of $V''(p^*)$ given in (\ref{stabi1}) tells us that it is a maximum of the potential.  Secondly by solving (\ref{com_fix}), we obtain another steady state and it is named as FP$_1$ in Fig. \ref{fig1} and the value of $V''(p^*)$ corresponding to this steady state confirms that it is a minimum of the potential $V(p)$.  Fig. \ref{fig1} also clearly shows that FP$_1$ is a minimum of the potential. Similarly, considering the turning points of the system, we find that one of the turning points is the same as the steady state FP$_2$ and the other turning points are obtained by solving Eq. (\ref{tpp}).  We find that there exists only one real positive valued turning point and it is denoted by TP in Fig. \ref{fig1}. 
\par  As mentioned earlier, the initial position of the particle and its initial velocity cannot be taken in a random way and it is fixed from the initial conditions of the coupled waveguide problem.  For example, in the case considered in Fig. \ref{fig1} the initial position is $p(0)$ ($p(0)=|\phi_1(0)|^2+|\phi_2(0)|^2=1.0$) and is shown by a filled red circle in Fig. \ref{fig1}. The initial velocity of the particle is $\dot{p}(0)=-(\gamma-\alpha p(0)) s_0(0)<0$ where $s_0(0)=|\phi_1(0)|^2-|\phi_2(0)|^2$ (vide Eq. (\ref{pd})).  Due to $\dot{p}(0)<0$, we find the value of $p$ decreases with respect to $z$ and so we observe that the particle in the potential travels towards left as shown by the arrows in Fig. \ref{fig1}.  While moving to the left, the particle reaches the turning point TP and at the point the velocity of the particle becomes zero but the acceleration of the particle is non-zero.  Such non-zero acceleration acclerates the particle in the opposite direction so that the value of $p$ starts to increase and the particle travels towards right.  The particle travels in the direction as shown by arrows in Fig. \ref{fig1}.  The next turning point that the particle encounters is FP$_2$ and as soon it reaches the point, the veloche ity of the particle becomes zero.  Now at this turning point, the particle is not accelerated to change the direction of its motion as the acceleration becomes zero at this point (as FP$_2$ is a fixed point so that $\ddot{p}|_{p=p^*}=-V'(p^*)=0$).  Due to the above reason, the particle settles down at the point FP$_2$ as soon it reaches that point.  Thus at $z \rightarrow \infty$, $p$ reaches the value $\frac{\gamma}{\alpha}$.   Note that this situation represents the stabilization of asymmetric modes (\ref{as_mod1}) in the coupled waveguide system where $p=|\phi_1|^2+|\phi_2|^2=\frac{\gamma}{\alpha}$.  Thus this particle picture helps to understand the dynamics of the coupled waveguide system satisfactorily. 
\subsubsection{Existence of blow-up responses and the role of $\chi$}
\par{  In the complete region of $k<\frac{\gamma(\alpha+\chi)}{2 \alpha}$ (the stable region of asymmetric mode (\ref{as_mod1})), we observe similar dynamics as above for certain range of initial conditions.  However for other initial conditions we observe blow-up solutions. To establish this, we first demonstrate the existence of blow-up solutions for high input powers and for the same parametric values as taken in Fig. \ref{fig1}. Note that the stability analysis discussed in Sec. III is a local analysis, and so it can predict the stable nature of the states only for initial conditions near the limit cycle attractor/asymmetric mode.  But for initial conditions away from the considered stationary states (modes), the stability analysis discussed in Sec. III may not provide information on whether the system will reach the stationary state (mode or attractor) as $z \rightarrow \infty$.  Thus the existence of blow-up solutions for high input powers are not so visible from Sec. III, but with this potential picture, we can capture the nature of the same. }
\par For instance, we have plotted the potential function for the {same set of parametric values as in Fig. \ref{fig1} and for different initial conditions}, namely $\phi_1(0)=\sqrt{2.5}+0.0i$ and $\phi_2(0)=0.0$ in Fig. \ref{fig2}(a) (as the values of $I_1$ and $I_2$ in $\widetilde{V}(p)$ changes with respect to the initial conditions $\phi_1(0)$ and $\phi_2(0)$, the form of $\widetilde{V}(p)$ also changes with respect to the initial conditions).  While trying to figure out the steady states of the potential, we find that { $\frac{\gamma}{\alpha}$ is the only steady state} of the potential and Eq. (\ref{fpp1}) does not provide any real roots.  Similarly there are {no other turning points} other than $\frac{\gamma}{\alpha}$.  Considering the steady state $p^*=\frac{\gamma}{\alpha}$, the value of $V''(p^*)$ given in Eq. (\ref{stabi1}) tells us that it is a maximum of the potential and it is also obvious from Fig. \ref{fig2}(a).  The initial position of the particle in the potential is shown in Fig. \ref{fig2}(a) as $p(0)$.  As the potential has no minimum, the strucutre of the potential makes the particle to slide down to infinity. Thus the value of $p \rightarrow \infty$ as $z \rightarrow \infty$.  This establishes the existence of blow-up solutions for this initial condition.   Note that the paramters have been assigned to take the same values as that of the ones considered in Fig. \ref{fig1}.  Thus this analysis clearly establishes that in the region $k<\frac{\gamma(\alpha+\chi)}{2 \alpha}$, for certain initial conditions the system tends towards the state $p=\frac{\gamma}{\alpha}$ (stabilization of asymmetric mode) as illustrated by Fig. \ref{fig1} and also there are initial conditions for which the value of $p$ can get blown up.

\par So far we have set $\chi=0.0$ in Figs. \ref{fig1} and \ref{fig2}(a).  Now we study the role of the nonlinear $\cal{PT}$ symmetric coupling by introducing $\chi$ and show that important changes occur in the structure of the potential in Fig. \ref{fig2}(b).  In Fig. \ref{fig2}(b), we have plotted the potential for three values of $\chi$, namely $\chi=0.0,0.3$ and $0.4$.  We find that the potential structure of $\chi=0.3$ looks qualitatively the same as $\chi=0.0$ but it changes dramatically in the case $\chi=0.4$.  For $\chi=0.4$, we find that the potential has two steady states, one at FP$_1$ and the other at FP$_2$, as shown in Fig. \ref{fig2}(b).  The system has two turning points one at TP and the other is same as FP$_2$.  The initial position of the particle $p(0)=2.5$ (as $\phi_1(0)=\sqrt{2.5}$ and $\phi_2(0)=0.0$) is indicated by a green square in Fig. \ref{fig2}(b) and the initial velocity $\dot{p}(0)>0$.   Thus the value of $p$ starts to increase and the particle moves towards the right and reaches the point TP.  At TP, the direction of motion of the particle is changed so that the particle starts to travel towards left and reaches the other turning point FP$_2$ at $\frac{\gamma}{\alpha}$.  As the latter is also a fixed point, the particle settles down at this point.  Thus at $z \rightarrow \infty$, $p \rightarrow \frac{\gamma}{\alpha}$ and it signifies that the coupled waveguide system tends towards the asymmetric mode (\ref{as_mod1}).   Thus $\chi$ provides control over the blow-up solutions and makes the asymmetric mode to become stable for a wide range of input powers.
\begin{figure} 
\includegraphics[width=1.05\linewidth]{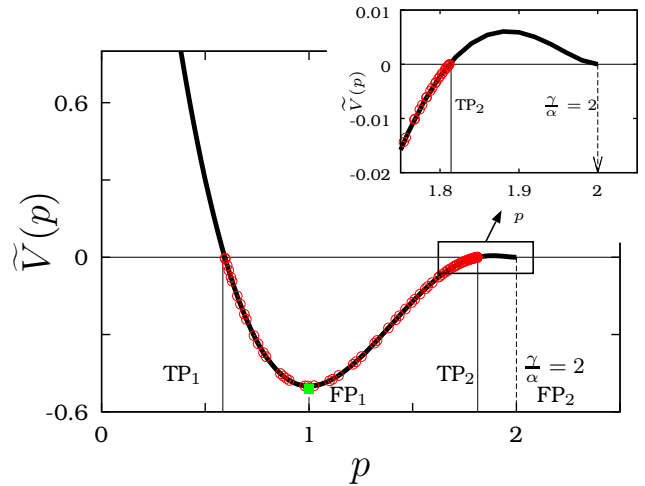}
\caption{(color online) Dynamics in the unbroken $\cal{PT}$ symmetric region for $\chi=0.0$: The figure shows the curves of $\widetilde{V}(p)$ for  $\gamma=1.0$, $\alpha=0.5$, $k=0.8$, $\chi=0.0$, $\phi_1(0)=1.0+0.0 i$ and $\phi_2(0)=0.0$. The dynamics in this region is the same as the dynamics observed in unbroken $\cal{PT}$ region of the coupled waveguide problem.}
\label{fig3}
\end{figure} 
\subsection{Dynamics in the unbroken $\cal{PT}$ region} 
\par So far, we have discussed the dynamics of the system for different initial conditions in the paramteric range $k<\frac{\gamma(\alpha+\chi)}{2 \alpha}$.   Now let us see how the system behaves and the potential function looks like in the region $k>\frac{\gamma(\alpha+\chi)}{2 \alpha}$.  Note that Eq. (\ref{stabi1}) indicates that in the latter case, $\frac{\gamma}{\alpha}$ becomes a minimum of the potential. 

\par By considering $k=0.8$ and $\chi=0.0$, we have plotted the potential function in Fig. \ref{fig3}.  For initial conditions $\phi_1(0)=1.0$ and $\phi_2(0)=0.0$, we observe that the potential has a minimum at FP$_1$ and  at FP$_2$.  In contrast to the previous case (corresponding to Fig. \ref{fig1}), Fig. \ref{fig3} shows that the potential has two turning points TP$_1$ and TP$_2$. The initial position of the particle is shown by a filled green square in Fig. \ref{fig3} and for the considered initial condition $\dot{p}(0)<0$ so that the value of $p$ decreases first and the particle moves towards left and reaches the turning point TP$_1$ and as soon it reaches TP$_1$, the direction of motion of the particle changes and, travels towards the rightside of the potential (or $p$ starts to increase). While travelling towards the rightside of the potential, the particle reaches the turning point TP$_2$.   As TP$_2$ is only the turning point and not a fixed point, the particle is acclerated to change its direction.  Thus we observe that the particle oscillates between the two turning points TP$_1$ and TP$_2$.   Due to the existence of TP$_2$ before FP$_2$, the particle never reaches the minimum at FP$_2$. The inset in Fig. \ref{fig3} shows the existence of a minimum at $\frac{\gamma}{\alpha}$.  Due to the presence of the new turning point, the system will not approach the point $\frac{\gamma}{\alpha}$ in the asymptotic limit, rather it executes oscillations between the two turning points TP$_1$ and TP$_2$.  This dynamics is equivalent to the stabilization of symmetric mode in the coupled waveguide problem where one observes oscillations in total power $p(z)$.  The values of the turning points exactly match with the maximum and minimum values of the oscillations observed in the total power of the system (\ref{syst}).  
\begin{figure} 
\includegraphics[width=1.05\linewidth]{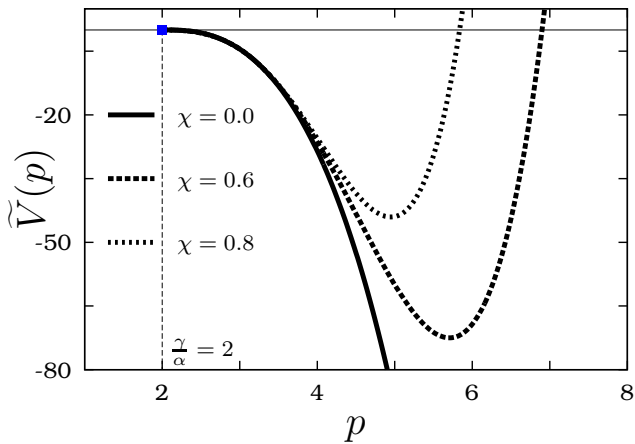}
\caption{(color online) Effect of nonlinear $\cal{PT}$ symmetric coupling in the unbroken $\cal{PT}$ region: The figure shows the curves of $\widetilde{V}(p)$ for  $\gamma=1.0$, $\alpha=0.5$, $k=0.8$, $\chi=0.0$, $0.6$, and $0.8$ where $\phi_1(0)=\sqrt{3.0}+0.0 i$ and $\phi_2(0)=0.0$. }
\label{fig4}
\end{figure}
\par Also we have noted earlier that even for this value of $k=0.8$, that is in the stable region of the symmetric modes, the system has blow-up solutions for certain input power configurations.  For example, one can consider the input power configuration for coupled waveguide system as $|\phi_1(0)|^2=3.0$ and $|\phi_2(0)|^2=0.0$ in such a case and the form of potential curve for $\chi=0.0$ is shown in Fig. \ref{fig4}.  From the figure, we can find that the considered system for this initial condition shows blow-up response.  Now with the introduction of $\chi$, by taking $\chi=0.6$ or $0.8$, we find that the structure of the potential is qualitatively changed in Fig. \ref{fig4}.  For these cases, we find that the particle tends to a steady state.  Because, for the choices $\chi=0.6$ or $0.8$, the potential does not look like the one given in Fig. \ref{fig3} whereas it looks more like the one given in Fig. \ref{fig1}, so that at the asymptotic limit ($z \rightarrow \infty$ ) the particle does not show oscillatory motion, rather it settles at the point $p=\frac{\gamma}{\alpha}$.  As mentioned earlier, this type of particle settlement at $p=\frac{\gamma}{\alpha}$ is similar to the stabilization of the symmetric mode. This is because that due to the increase in $\chi$, the value of $\frac{\gamma(\alpha+\chi)}{2 \alpha}$ increases so that  $k<\frac{\gamma(\alpha+\chi)}{2 \alpha}$ which leads to the stabilization of asymmetric mode.  This also denotes that the nonlinear coupling not only suppresses the blow-up regions but also favours or widens the symmetry broken region. 

\section{\label{mis} Asymmetric case}
\par As mentioned in Sec. \ref{modeel}, the above $\cal{PT}$ symmetric case can be regarded as a limiting case of a practically achievable asymmetric case mentioned in Eq. (\ref{syst_0}).  However the earlier $\cal{PT}$ symmetric case helps us to clearly visualize the usefulness of the nonlinear coupling in localizing light in a waveguide and in controlling blow-up responses.  The above behaviors of nonlinear coupling can be seen even in this asymmetric case as well.  In this asymmetric case, one has to consider $\omega_1>\omega_2$ in Eq. (\ref{syst_0}).  Although one can find a few qualititative changes in the dynamics of the system for $\omega_1 \neq \omega_2$, that is the region of power oscillations is largely suppressed in this case, here also we can observe stabilization of asymmetric modes which enable unidirectional light transport.  Even in this case, the nonlinear coupling is useful in localizing light in a waveguide and in controlling blow-up responses.  For example, to show the ability of the nonlinear coupling in localizing light in one of the waveguides, we have plotted Fig. \ref{mis1}.  From the figure, we can find that in the absence of nonlinear coupling $\chi$, the power is maximally localized in the second waveguide but the power in the first waveguide is not minimized.  Now with the introduction of nonlinear coupling as mentioned in Sec. \ref{modeel}, the stimulated Raman scattering induces energy transfer from pump mode to  Stokes mode.  Due to this reason, we observe that the intensity of light in the first waveguide is further minimized for $\chi=0.3$ as given in Fig. \ref{mis1} and it shows the ability to transport light unidirectionally (the light can be further localized in the second waveguide if one considers self-trapping nonlinearity into account).  Secondly, to show the role of nonlinear coupling in controlling blow-up responses in this case, we have plotted Figs. \ref{mis2}(a) and \ref{mis2}(b).  From Fig. \ref{mis2}(a), we find the existence of a wide blow-up region.  By the introduction of $\chi$, we can observe a large reduction in the blow-up region as shown in Fig. \ref{mis2}(b).  Thus it is clear from the above discussion that this type of Raman coupling is useful in controlling blow-up responses and to achieve unidirectional transport of light. 
\begin{figure} 
\includegraphics[width=0.7\linewidth]{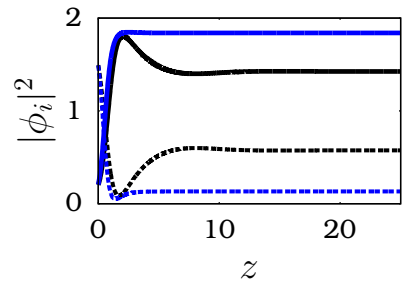}
\caption{(color online) The ability of nonlinear coupling in localizing light in the case $\omega_1 \neq \omega_2$. The figure is plotted for two values of $\chi$, the black curves correspond to the case $\chi=0.0$ and the blue (gray) curves corresponds to the case $\chi=0.3$ with $\omega_1=1.0$ and $\omega_2=0.8$.  Note that in Eq. (\ref{syst_0}), $\chi_1=\frac{\omega_1}{\omega_2} \chi$ and $\chi_2=\chi$.  In the figure, the curve corresponding to $|\phi_1|^2$ is represented by dotted lines and the curve corresponding to $|\phi_2|^2$ is represented by continuous lines. The figure is plotted for $\gamma=1.0$, $\alpha=0.5$ and $k=0.5$.}
\label{mis1}
\end{figure}
\begin{figure}[ht] 
\includegraphics[width=1.05\linewidth]{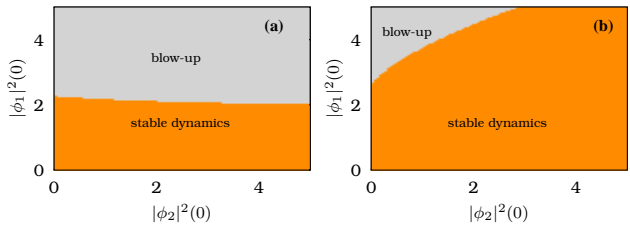}
\caption{(color online)Role of nonlinear coupling in controlling blow-up responses. Fig. (a) is plotted for $\chi=0.0$ and (b) is for $\chi=0.3$ and in both the figures, we considered $\gamma=1.0$, $\alpha=0.5$, $\omega_1=1.0$, $\omega_2=0.8$ and $k=0.5$ (the same parametric values as chosen in Fig.\ref{mis1}). We here varied the initial conditions and pictured out the input power configurations that lead to blow-up.  To draw Figs. (a) and (b), we considered $\phi_1(0)=u(0)+0i$ and $\phi_2(0)=v(0)+0i$. }
\label{mis2}
\end{figure}
\section{Summary}
\par In this article, we have studied the role of an interesting nonlinear $\cal{PT}$ symmetric coupling in a coupled waveguide system and importantly we looked onto the problem of achieving stable unidirectional light transport.  As one can see in \cite{srep}, achieving such stable unidirectional transport of light is more essential from the application point of view.  We here considered a system with saturating loss-gain profile which can enable transportation of stable finite power asymmetric modes.  The stable nature of the asymmetric and symmetric modes in an interesting $\cal{PT}$ symmetric case has been illustrated.  We have also mentioned that there are two drawbacks in the system in achieving unidirectional light transport (i) in localizing light completely in one of the waveguides and (ii) in controlling blow-up responses.  With an integrable $\cal{PT}$ symmetric case, we have clearly illustrated that these two demerits can be overcome by a nonlinear coupling that is relevant to the stimulated Raman scattering.   However, this $\cal{PT}$ symmetric case can be regarded as a limiting case of the practically realizable asymmetric case.  Even in this asymmetric case, we can achieve unidirectional light transport and control blow-up responses which are explained in Sec. VII.  We hope that the results on the ability to have controlled finite power unidirectional light transport and to control blow-up responses can find fruitful applications in the construction of unidirectional optical elements such as optical diodes and optical isolators.   Efforts to engineer such models can reveal their adapatability and applicability.

\section*{Acknowledgement}
SK thanks the Department of Science and Technology (DST), Government of India, for providing a INSPIRE Fellowship.  The work of VKC is supported by the SERB-DST Fast Track scheme for young scientists under Grant No.YSS/2014/000175.  The work of MS forms part of a research project sponsored by Department of Science and Technology, Government of India.  The work of ML is supported by a NASI Senior Scientist Platinum Jubilee fellowship program.

\end{document}